\documentclass[graybox]{svmult}

\usepackage{type1cm}        
\usepackage{makeidx}         
\usepackage{graphicx}        
\usepackage{multicol}        
\usepackage[bottom]{footmisc}

\usepackage{newtxtext}       %
\usepackage[varvw]{newtxmath}       

\makeindex             

\begin{document}

\title*{Recovering the CMB signal with neural networks}
\author{José Manuel Casas\orcidID{0000-0001-7882-1691},\\ Laura Bonavera\orcidID{0000-0001-8039-3876},\\ Joaquín González-Nuevo\orcidID{0000-0003-1354-6822},\\ Giuseppe Puglisi\orcidID{0000-0002-0689-4290} and \\ Carlo Baccigalupi\orcidID{0000-0002-8211-1630}}
\institute{José Manuel Casas \at Instituto de Ciencias y Tecnologías Espaciales de Asturias, Universidad de Oviedo, C. Federico Garc{\'i}a Lorca 18, 33007 Oviedo, Spain, \email{casasjm@uniovi.es}}
%
%
\maketitle


\abstract{Component separation is the process of extracting one or more emission sources in astrophysical maps. It is therefore crucial to develop models that can accurately clean the cosmic microwave background (CMB) in current and future experiments. In this work, we present a new methodology based on neural networks which operates on realistic temperature and polarization simulations. We assess its performance by comparing the power spectra of the output maps with those of the input maps and other emissions. For temperature, we obtain residuals of 20$\pm$10\hspace{1pt}$\mu K^{2}$. For polarization, we analyze the $E$ and $B$ modes, which are related to density (scalar) and primordial gravitational waves (tensorial) perturbations occurring in the first second of the Universe, obtaining residuals of $10^{-2}$\hspace{1pt}$\mu K\hspace{1pt}^{2}$ at $l>200$ and $10^{-2}$ and $10^{-3}$\hspace{1pt}$\mu K\hspace{1pt}^{2}$ for $E$ and $B$, respectively. 
}
\section{Introduction}
\label{sec:1}

The cosmic microwave background (CMB) corresponds to the afterglow of the Big Bang, dating back to about 380000 years after the birth of the Universe, when photons decoupled from baryons and the Universe became transparent to light \cite{WEI08}. It exhibits small deviations in temperature and polarization that can be described by the T, Q, and U Stokes parameters. The CMB polarization can also be decomposed into gradient and curl modes, known as the $E$ and $B$ modes, respectively \cite{ZAL97}.

While the temperature power spectra and the $E$ mode polarization and their correlation have been successfully reconstructed in years, particularly by the Planck satellite \cite{PLA_2018_I}, the $B$ mode polarization, both cosmological and lensed, remain much more challenging to detect, as it is expected to be fainter than the $E$ modes. 

Ongoing research therefore is focused on improving not only the sensitivity of future experiments but also developing new techniques to clean the CMB from other physical emissions in the microwave sky with the use of multi-frequency observations and simulations. In this work, we present a new component separation method based on neural networks, and it is called the cosmic microwave background extraction neural network (CENN). It was firstly presented in \cite{CAS22b} for recovering CMB temperature maps and later evaluated on polarization in \cite{CAS24}, both in realistic simulations of the Planck experiment. In this work, we summarize the main results from both analyses.

\section{Simulations}
\label{sec:simulations}

Our data consist of realistic simulations of the microwave T, Q, and U microwave maps as observed by \textit{Planck} at 143, 217, and 353 GHz in the case of temperature and at 100, 143, and 217 GHz for polarization. The maps were downloaded from the Planck Legacy Archive website\footnote{http://pla.esac.esa.int/pla/\#home.}. They were cut using the methodology described in \cite{KRA21} into square patches of 256$\times$256 pixels, with a pixel size of 90 arcseconds. 

Each patch includes the CMB signal, random white instrumental noise at \textit{Planck} levels, Galactic (interstellar dust and synchrotron emission) and extragalactic foregrounds (Sunyaez-Zeldovich effects, the Cosmic Infrared Background and radio point sources) \cite{DEL12}.  Thus, on each frequency, we generate a patch at the same sky location for each component, then filter each one with the full width at half maximum (FWHM) of the instrument. These components are then summed, and instrumental noise is added at the corresponding Planck levels. Finally, we simulate three types of datasets: one for training, one for validating the network during the training, and one for testing the model in the final analysis. 

\section{Methodology}
\label{sec:methodology}

\begin{figure*}
\centering
\includegraphics[width=\linewidth]{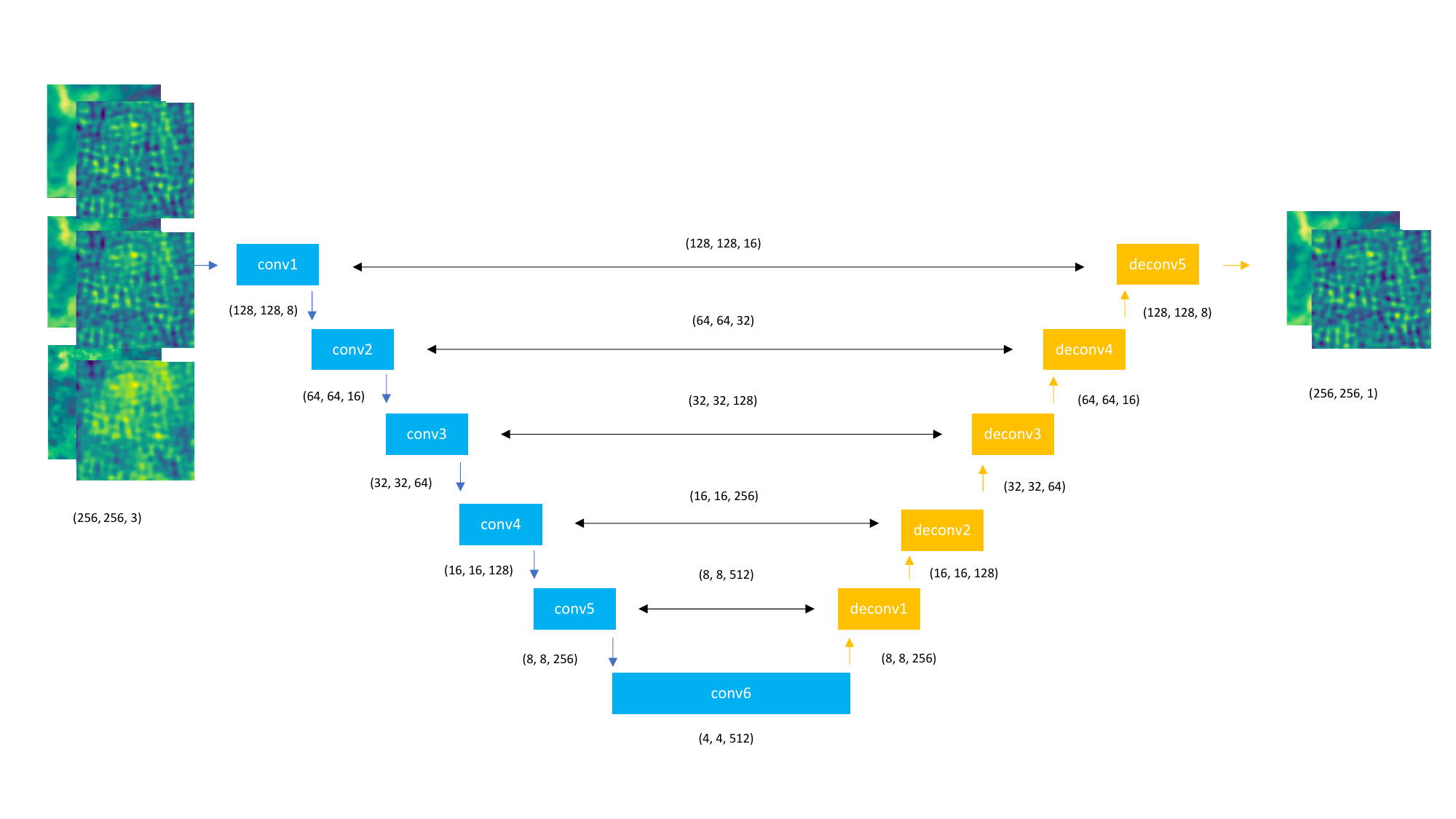}
\caption{Architecture of CENN, the neural network used in this work. We show for visualization purposes the kind of input and output patches in both temperature (back) and polarization (front).}
\label{Fig:Architecture}
\end{figure*}

The details of the neural network architecture used in this work, and how it performs image segmentation on multi-frequency microwave data, are explained in \cite{CAS22b}. The reader is encouraged to consult that work; however, we also provide here a brief summary of how the network functions.

The neural network is trained to take as input a set of three patches of the microwave sky and to output a patch containing only the CMB signal at the central frequency channel. In the first convolutional block, the network reads the three input patches, and the information is processed through 3-channel filters that operate across the three frequencies for each of the initial feature maps. The output from the activation function is then passed through subsequent convolutional blocks.
Deconvolutional blocks, connected in sequence to the convolutional ones, are used to extract the CMB signal from the intermediate feature maps. Finally, the last deconvolutional block reconstructs the CMB signal into a patch with the same dimensions as the input patches.

\section{Results}
\label{sec:results}

We have divided our work into CMB recovering in temperature, summarized in Section \ref{sec:temperature}, and its extension to polarization simulations, shown in Section \ref{sec:polarization}. 

\subsection{Temperature}
\label{sec:temperature}

\begin{figure*}[ht]
\centering
\includegraphics[width=0.5\linewidth]{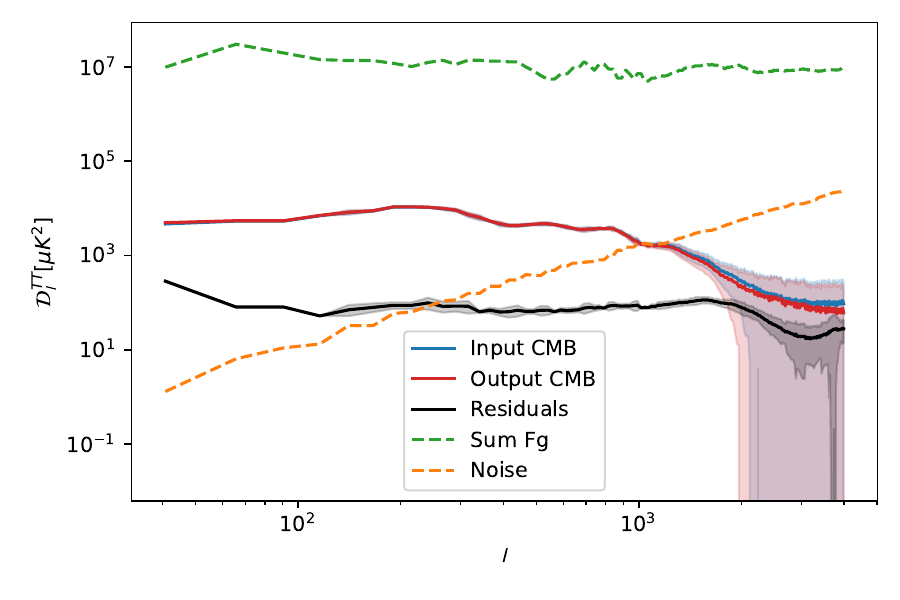}
\caption{Comparison between the input CMB in the validation simulations (in blue) with respect the output from the neural network (in red) and the residuals (in black). Green and orange lines show the level of the contaminants. Shaded areas represent the 1$\sigma$ standard deviation for each bin \cite{CAS22b}.}
\label{Fig:results_temperature}
\end{figure*}

In this subsection, we analyze the average power spectrum of the recovered CMB compared to the input in the validation simulations, as well as the sum of the foregrounds. The results are summarized in Figure \ref{Fig:results_temperature}.

We obtain a difference between input (in blue) and output CMB (in red) of 13$\pm$110\hspace{1pt}$\mu K^{2}$, with residuals, computed as the power spectra of that difference, of approximately 20$\pm$10\hspace{1pt}$\mu K^{2}$. These are promising results when compared with the sum of the foregrounds (in green) and the instrumental noise (in orange). The foregrounds dominate the signal at large and medium scales, while the instrumental noise is dominant at small scales.

For comparison, traditional methods used in Planck (SMICA, NILC, Commander, and SEVEM\footnote{These methods are described in detail in \cite{PLA_2015_IX}.}), validated in realistic simulations in \cite{PLA_2015_IX}, estimated the power spectra with a mean absolute error of 15$\pm$100\hspace{1pt}$\mu K^{2}$. These results are generally comparable to those obtained by the neural network, except at small scales, where the network continues to estimate the spectra above $l\sim1000$,  whereas traditional methods begin to overestimate the signal due to instrumental noise.

\subsection{Polarization}
\label{sec:polarization}

\begin{figure*}
\minipage{0.5\textwidth}
\includegraphics[width=\linewidth]{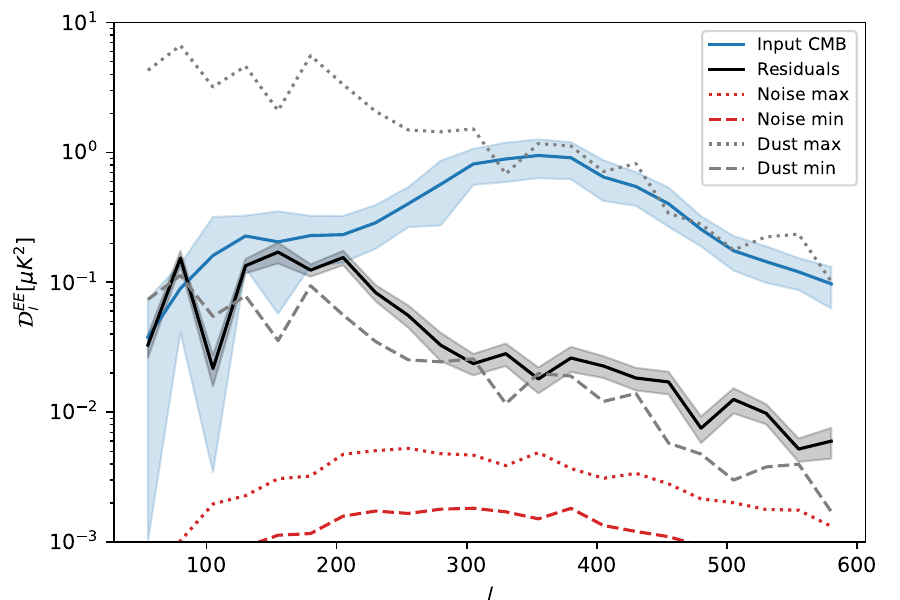}
\endminipage
\minipage{0.5\textwidth}%
  \includegraphics[width=\linewidth]{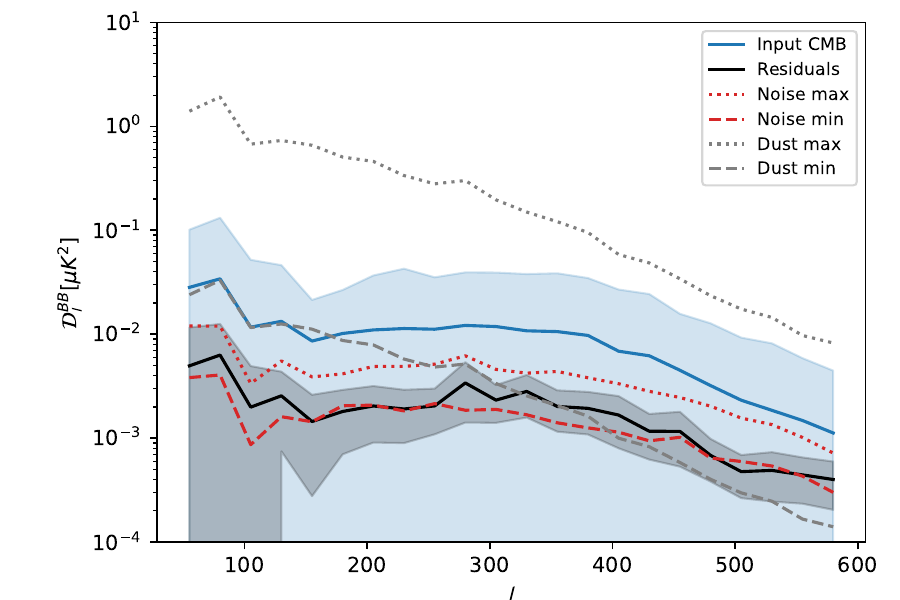}
\endminipage
\caption{Comparison between the input CMB in the validation simulations (in blue) with respect residuals (in black). Grey and red lines show the level of the contaminants. Shaded areas represent the 1$\sigma$ standard deviation for each bin \cite{CAS24}.}
\label{Fig:results_polarization}
\end{figure*}

In this section, we follow a similar analysis than in the previous one. In this case, we combine Q and U maps to analyze both $E$ and $B$ modes. These results are summarized in Figure \ref{Fig:results_polarization}.

As shown in the left panel, the network recovers the $E$ mode with residuals generally lower than the input CMB, from $10^{-1}$\hspace{1pt}$\mu K\hspace{1pt}^{2}$ at large scales, decreasing to $10^{-2}$\hspace{1pt}$\mu K\hspace{1pt}^{2}$ at medium scales. At smaller scales, residuals are below $10^{-2}$\hspace{1pt}$\mu K\hspace{1pt}^{2}$. Attending to dust levels (in grey), large scale residuals seem to be artifacts generated by the network when recovering the CMB due to Galactic contamination. For the $B$ mode, we obtain $2 \times 10^{-3}$\hspace{1pt}$\mu K\hspace{1pt}^{2}$ for $l < 400$, decreasing to $5 \times 10^{-4}$\hspace{1pt}$\mu K\hspace{1pt}^{2}$ at $l > 500$.

For the $E$ mode, we obtain similar results than the traditional methods cited in the previous section, since they estimated the spectra with a mean absolute difference between input and output signal between 0 and 1 $\mu K\hspace{1pt}^{2}$, and a standard deviation between 5 and 6 $\mu K\hspace{1pt}^{2}$ in realistic Planck simulations at 10' FWHM \cite{PLA_2015_IX}, as so as the spectra was accurately recovered up to l$\sim$700, where noise started to dominate the signal. For the $B$ mode, traditional methods could not estimate the signal nor give an upper limit and we cannot directly compare with them, at least for this kind of data. With the neural network it seems that, at these noise and foreground levels, it would give an upper limit on Planck data, which will be further investigated in a future work.

\section{Conclusions}
\label{sec:conclusions}

Component separation is the process of extracting one or more astrophysical signals from simulations and observations. This is crucial for future cosmic microwave background (CMB) experiments, which will be designed to search for primordial gravitational waves from the Cosmic Inflation period, imprinted on the CMB. In this work, we present a new methodology based on neural networks. It summarizes the performance of recovering temperature CMB maps in Planck realistic temperature simulations \cite{CAS22b} and its extension to polarization maps in \cite{CAS24}.

In both cases, we train and validate the network using realistic simulations of \textit{Planck} in the form of sky quadratic patches. We then analyze its performance by comparing the mean power spectrum of all the validation patches with respect to the impact of other emissions, called foregrounds.

For temperature maps, we obtain a difference between input and output CMB of 13 $\pm$ 110 $\mu K^{2}$ with residuals of 20 $\pm$ 10 $\mu K^{2}$, and with the average level of contamination by foregrounds being around 3 orders of magnitude above the signal. For polarization maps, we recover the $E$ mode with residuals of $10^{-2}$\hspace{1pt}$\mu K\hspace{1pt}^{2}$ and of $2 \times 10^{-3}$\hspace{1pt}$\mu K\hspace{1pt}^{2}$ for the $B$ mode.

In both analyses, we found similar estimates to traditional methods when applied to Planck realistic simulations. However, it seems that the neural network separates the signal with higher accuracy at small scales, when instrumental noise starts to dominate the maps. The evolution of the current CENN approach in polarization should involve using larger patches in order to reach lower multipoles and provide estimates for the tensor-to-scalar ratio. This is now being considered in a follow-up work. Additionally, a similar network could be used to constrain the properties of polarized foregrounds, which is another project that is nearly ready for submission.

\begin{acknowledgement}
The authors warmly thank both referees for their commentaries and suggestions that improve this work. JMC, LB and JGN acknowledge the CNS2022-135748 project funded by MCIN/AEI/10.13039/501100011033 and the PID2021-125630NB-I00 project funded by MCIN/AEI/10.13039/501100011033/FEDER, UE. JMC also acknowledges the SV-PA-21-AYUD/2021/51301 grant. GP acknowledges support from Italian Research Center on High Performance Computing Big Data and Quantum Computing (ICSC), project funded by European Union NextGenerationEU and National Recovery and Resilience Plan (NRRP) Mission 4 Component 2 within the activities of Spoke 3 (Astrophysics and Cosmos Observations). CB acknowledges support from the COSMOS project of the Italian Space Agency, and the INDARK Initiative of the INFN.
\end{acknowledgement}
\ethics{Competing Interests}{The authors have no conflicts of interest to declare that are relevant to the content of this chapter.}

\end{document}